\documentclass[preprint2]{aastex6}
\usepackage{rotating}
\usepackage{graphicx}
\usepackage{amssymb}
 \renewcommand{\thefootnote}{\fnsymbol{footnote}}

\shortauthors{French et al.}
\slugcomment{Draft \date}
\shorttitle{Why Post-Starburst Galaxies Are Now Quiescent}

\newcommand\um      {\ifmmode\mu{\rm m}\else$\mu${\rm m}\fi\xspace}

\begin{document}

\title{Why Post-Starburst Galaxies Are Now Quiescent}

\author{K. Decker French$^\dagger$}
\affil{Observatories of the Carnegie Institute for Science, 813 Santa Barbara Street, Pasadena CA 91101 \\ Steward Observatory, University of Arizona, 933 North Cherry Avenue, Tucson AZ 85721}
\and
\author{Ann I. Zabludoff}
\affil{Steward Observatory, University of Arizona, 933 North Cherry Avenue, Tucson AZ 85721}

\author{Ilsang Yoon}
\affil{National Radio Astronomy Observatory, 520 Edgemont Road, Charlottesville, VA 22903}

\author{Yancy Shirley}
\affil{Steward Observatory, University of Arizona, 933 North Cherry Avenue, Tucson AZ 85721}

\author{Yujin Yang}
\affil{Korea Astronomy and Space Science Institute, 776 Daedeokdae-ro, Yuseong-gu, Daejeon 305-348, Korea \\ Korea University of Science and Technology (UST), 217 Gajeong-ro Yuseong-gu, Daejeon 34113, Korea}

\author{Adam Smercina}
\affil{Department of Astronomy, University of Michigan, 1085 S. University Avenue, Ann Arbor, MI 48109}

\author{J.D. Smith}
\affil{Department of Physics and Astronomy, University of Toledo, Ritter Obs., MS \#113, Toledo, OH 43606}

\and
\author{Desika Narayanan}
\affil{Department of Astronomy, University of Florida, 211 Bryant Space Sciences Center, Gainesville, FL 32611 \\ Cosmic Dawn Center (DAWN), Niels Bohr Institute, University of Copenhagen, Juliane Maries vej 30, DK-2100 Copenhagen, Denmark}

\begin{abstract}

Post-starburst or ``E+A" galaxies are rapidly transitioning from star-forming to quiescence. While the current star formation rate of post-starbursts is already at the level of early type galaxies, we recently discovered that many have large CO-traced molecular gas reservoirs consistent with normal star forming galaxies. These observations raise the question of why these galaxies have such low star formation rates. Here we present an ALMA search for the denser gas traced by HCN (1--0) and HCO$^+$ (1--0) in two CO-luminous, quiescent post-starburst galaxies. Intriguingly, we fail to detect either molecule. The upper limits are consistent with the low star formation rates and with early-type galaxies. The HCN/CO luminosity ratio upper limits are low compared to star-forming and even many early type galaxies. This implied low dense gas mass fraction explains the low star formation rates relative to the CO-traced molecular gas and suggests the state of the gas in post-starburst galaxies is unusual, with some mechanism inhibiting its collapse to denser states. We conclude that post-starbursts galaxies are now quiescent because little dense gas is available, in contrast to the significant CO-traced lower density gas reservoirs that still remain.

\end{abstract}


%%%%%%%%%%%%%%%%%%%%%%%%%%%%%%%%%%%%%%%%%%%%%%%%%%%%%%%%%%%%%%%%%%%%%%%%%%%%%%%%%

\section{Introduction}

\footnotetext[0]{$^\dagger$ Hubble Fellow}
\renewcommand*{\thefootnote}{\arabic{footnote}}
\setcounter{footnote}{0}

The nature of how and when the molecular gas reservoirs are depleted in galaxies is essential to understanding the question of why galaxies become quiescent. To explore this question, we focus on a class of galaxies in the midst of rapid evolution in their star formation properties. The spectra of post-starburst, ``E+A", or ``k+a"  galaxies show strong Balmer absorption, indicative of a recent starburst that ended in the last Gyr, yet little ongoing star formation, indicating a rapid change from star-forming to quiescent \citep{Dressler1983,Couch1987}. Despite their low star formation rates (SFRs), many (over half of those studied) have large CO-traced molecular gas reservoirs \citep{French2015,Rowlands2015,Alatalo2016b}. These quiescent post-starburst galaxies have similar CO-traced molecular gas fractions (M$_{\rm mol}/$M$_\star$) as normal star-forming galaxies, implying a lower CO-traced star formation efficiency SFE $\propto$ SFR/$L^{\prime}($CO$)$. This offset persists in the classical Kennicutt-Schmidt \citep{Kennicutt1998} relation, suggesting post-starburst galaxies experience a $\sim 4 \times$ suppression of SFE in the CO-traced molecular gas.

Previous studies of the molecular gas content of post-starburst galaxies have used the CO (1--0) and CO (2--1) lines as tracers. The CO (1--0) line is sensitive to molecular gas at densities $\sim 100$ cm$^{-3}$. Other molecules, such as HCN (1--0) and HCO$^+$ (1--0), trace denser gas. The critical densities of HCN (1--0) and HCO$^+$ (1--0) are $3\times10^6$ cm$^{-3}$ and $2\times10^5$ cm$^{-3}$ \citep[e.g.,][]{Juneau2009}, although they are also sensitive to less dense gas, with effective excitation densities of $8.4\times10^3$ cm$^{-3}$ and $950$ cm$^{-3}$, respectively \citep{Shirley2015}. The HCN (1--0) luminosity correlates more linearly with SFR, and with less scatter, than does the CO (1--0) luminosity \citep{Gao2004}, even down to the scales of star-forming clumps \citep{Wu2005,Wu2010}. In starburst galaxies, which have high CO-traced SFEs, the ratio of HCN to CO luminosity is high, resulting in a high dense gas mass consistent with the high SFRs. An analogous situation may apply in the post-starburst galaxies: if we measure a low dense gas mass, this would be consistent with the low SFRs in these galaxies, despite the higher CO luminosities.

We must measure the properties of this denser gas in post-starburst galaxies to understand why there is no significant star formation and why CO-traced molecular gas remains. Here, we present an ALMA survey of HCN (1--0) and HCO$^+$ (1--0) in two CO-luminous post-starburst galaxies.

\section{Data}

\subsection{ALMA Observations}
\label{sec:almaobs}

\begin{figure*}
\centering
\includegraphics[width = 0.49\textwidth]{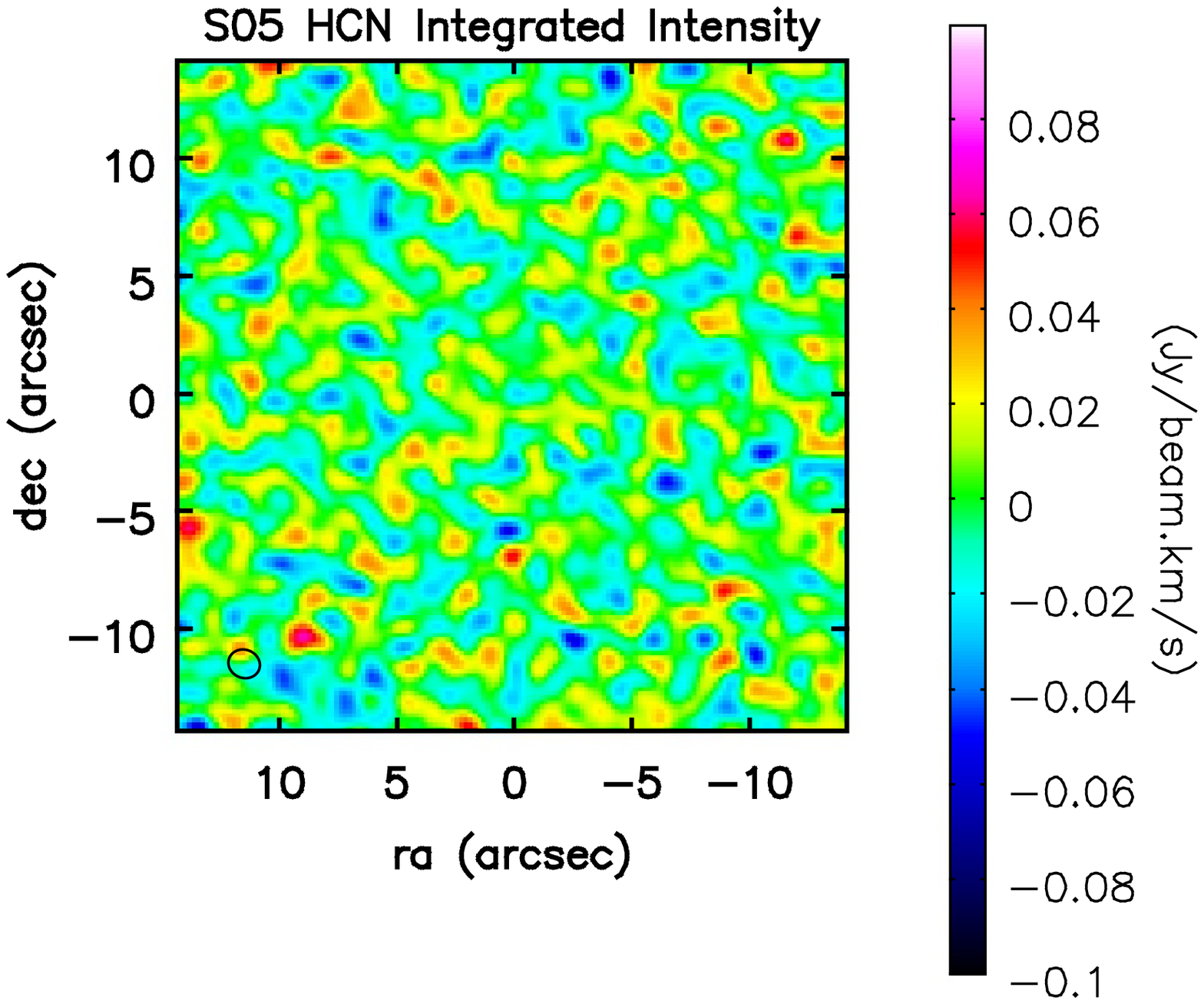}
\includegraphics[width = 0.49\textwidth]{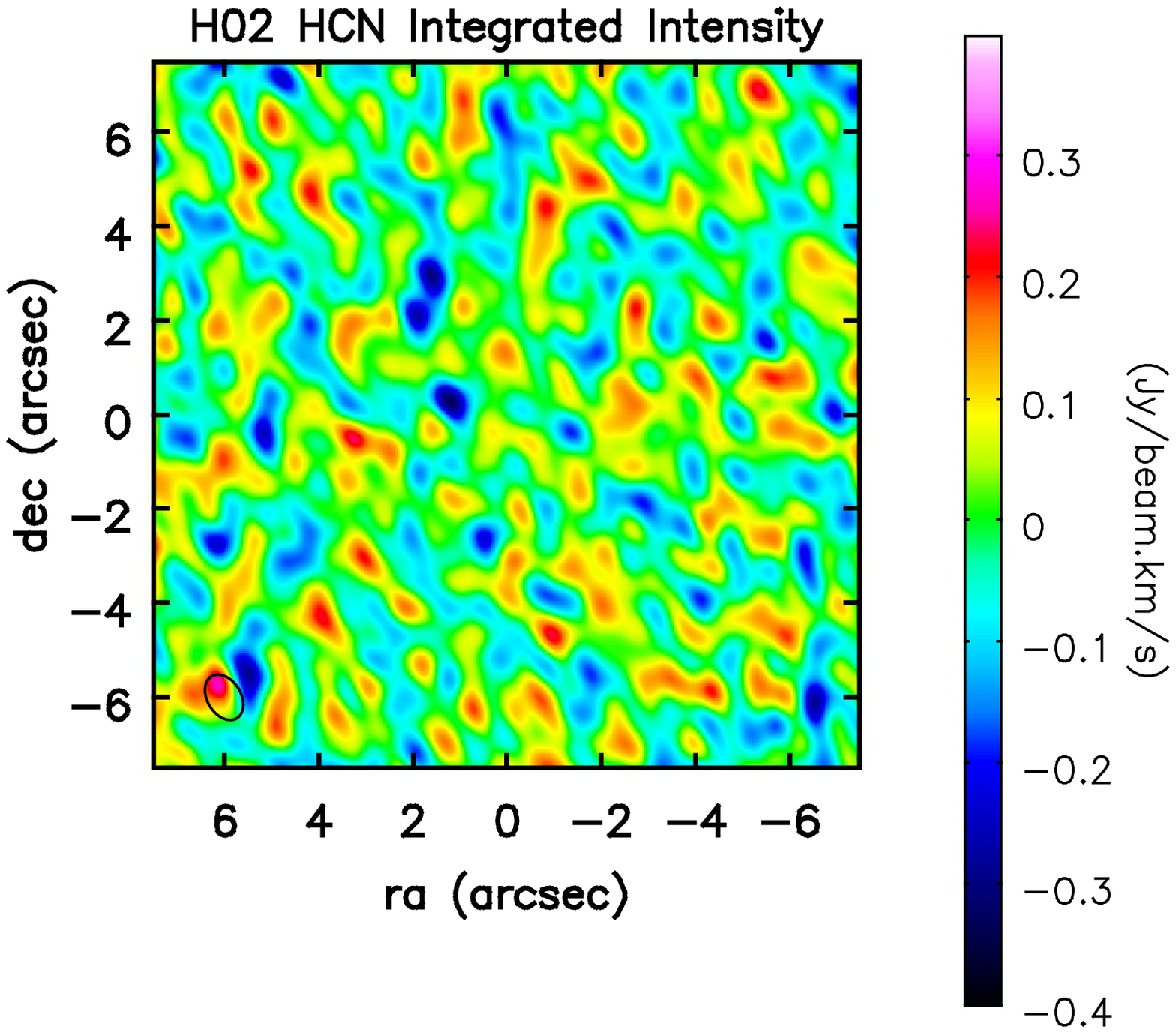}
\caption{Integrated intensity maps for the two post-starburst targets for HCN (1--0). The galaxy optical centers are at (0,0) on each plot. Neither source is detected.}
\label{fig:data}
\end{figure*}

The observations for this work were obtained during ALMA Cycle 4 (program 2016.1.00881.S; PI: French). We observe two post-starburst galaxies with CO (1--0) detections from \citet{French2015} with representative CO luminosities and SFRs, and with redshifts and declinations enabling observations with ALMA (labeled H02 and S05). We use the Band 3 receiver (84-116 GHz) with two spectral windows to observe HCN 1--0 (88.63 GHz rest frame), HCO$^+$ 1--0 (89.19 GHz rest frame), and HNC 1--0 (90.66 GHz rest frame). The redshift of H02 pushes HCN close to the edge of the Band 3 bandwidth, so we adopt narrow spectral windows, of 417, 831, 817 km/s for the lines, respectively. These spectral windows are still much larger than the 170 km/s linewidth of the CO (1--0) line for this galaxy. For S05, we adopt wider spectral windows of 1622-1660 km/s. 

The ALMA observations described here have a beam size of 1-1.5\arcsec\ (or 1.0-1.4 kpc), with a maximum recoverable scale of 7.6\arcsec. The beam size is chosen to match the observed CO (2--1) sizes, as our companion ALMA program (2015.1.00665.S; PI: Smith; Smercina et al. in prep) finds the CO (2--1) emission confined to the central 1-1.5\arcsec\ of these galaxies. Both the CO (2--1) sizes and Spitzer IRAC 8$\mu$m sizes are smaller than the optical sizes. The optical $r$-band half-light radii of these galaxies is 1.71\arcsec\ and 2.74\arcsec\ for H02 and S05, respectively. The resolved CO (2--1) fluxes are consistent with the unresolved IRAM 30m CO (2--1) observations (with $\sim$11\arcsec\ beamsize), indicating a lack of significant CO emitting gas beyond the central 1-2 kpc of these galaxies. Thus, it is unlikely that significant HCN emission is resolved out by our observations, as HCN emission is unlikely to be found outside of the CO emitting regions. 

Additionally, interferometric observations typically constrain HCN emission to come from the central kpc of galaxies \citep[e.g.,][]{Aalto2012,Kepley2013,Scoville2015, Chen2015}. Even when this dense gas is more extended or in outflows \citep[e.g.,][]{Alatalo2014,Salas2014}, the HCN emission does not extend more than 500 pc from the main disk. The ALMA HCN observations can therefore be directly compared to the single dish CO (1--0) measurements.

The observations of H02 were executed in one block on November 4, 2016. The observations of S05 were executed in four blocks on November 22 and 25, 2016, and May 4, 2017. The 12m array was used for both datasets, using 43 antennas for H02 and 41-46 antennas for S05. The observations were carried out in configurations C40-5 and C40-4 for H02 and S05, respectively.

The data were pipeline-calibrated using CASA version 4.7.2. The H02 observations of HCN (1--0) have a final beamsize of 1.0\arcsec$\times$0.73\arcsec\ and a sensitivity of 5.5 mJy/beam at a spectral resolution of 0.252 km/s. The H02 observations of HCO$^+$ (1--0) have a final beamsize of 1.1\arcsec$\times$0.72\arcsec\ and a sensitivity of 2.7 mJy/beam at a spectral resolution of 1.004 km/s. Neither line is detected. The S05 observations of HCN (1--0) and HCO$^+$ (1--0) are also non-detections. The S05 observations of HCN (1--0) have a final beamsize of 1.4\arcsec$\times$1.2\arcsec\ and a sensitivity of 570 $\mu$Jy/beam at a spectral resolution of 1.729 km/s. The S05 observations of HCO$^+$ (1--0) have a final beamsize of 1.3\arcsec$\times$1.2\arcsec\ and a sensitivity of 570 $\mu$Jy/beam at a spectral resolution of 1.718 km/s. In order to calculate upper limits on the integrated intensities, we assume the same linewidths as the CO observations: 172 km/s for H02 and 350 km/s for S05.

We test the robustness of these upper limits in several ways. First, we taper the data to 5\arcsec, but still do not detect either galaxy in HCN (1--0). Second, we use a matched filter technique \citep{Loomis2018} using the prior information from the CO observations, but do not find any significant detections just below our sensitivity limit.

\begin{deluxetable*}{lccccccc}
\tabletypesize{\scriptsize}
\tablewidth{0pt}
\tablecolumns{8}
\tablecaption{Post-Starburst ALMA Observations\label{table:hcn}}
\tablehead{\colhead{Name} & \colhead{RA (J2000)} & \colhead{Dec (J2000)} & \colhead{z} & \colhead{$L^{\prime}($CO$)$\tablenotemark{a}} & \colhead{SFR\tablenotemark{b}} & \colhead{$L^{\prime}($HCN$)$\tablenotemark{c}} & \colhead{$L^{\prime}($HCO$^+$$)$\tablenotemark{c}} \\
\colhead{} & \colhead{(deg)} & \colhead{(deg)} & \colhead{}  & \colhead{($10^6$ K\,km\,s$^{-1}$\,pc$^2$)} & \colhead{(M$_\sun$ yr$^{-1}$)} & \colhead{($10^6$ K\,km\,s$^{-1}$\,pc$^2$)} & \colhead{($10^6$ K\,km\,s$^{-1}$\,pc$^2$)}}
\startdata
H02 &      141.580383 &       18.678055 &          0.0541 &             842 $\pm$            174 &            $<1$ & $<           29.3$ &$<           28.1$ \\
S05 &      146.112335 &        4.499120 &          0.0467 &             304 $\pm$             95 &            0.58 & $<            8.9$ &$<            8.8$ \\

\enddata
\tablenotetext{a}{\citet{French2015}}
\tablenotetext{b}{From H$\alpha$ flux for S05 and TIR flux for H02 (see \S\ref{sec:sfr}).}
\tablenotetext{c}{$3\sigma$ upper limits}
\end{deluxetable*}

\subsection{Star Formation Rates}
\label{sec:sfr}

Determining SFRs for post-starburst galaxies is complicated by the recent starburst, possible AGN activity, and heating from the young A-star population. In \citet[\S2.5]{French2015}, we explored many possible methods for calculating SFRs in the post-starburst galaxies using the available archival optical and radio data. Recent work by \citet{Smercina2018} has explored a number of infrared tracers for these same galaxies. Here, we discuss the constraints on the SFRs of the two galaxies considered here from various measures.

In \citet{French2015}, we compared the SFRs derived from H$\alpha$ emission and the 4000\AA\ Balmer break D$_n(4000)$, corrected for dust based on the Balmer decrement and for aperture based on their SDSS colors. A decreased SFE in the CO-traced gas was seen for both measures, which is conservative because both SFR indicators are likely upper limits: the high incidence of LINER-like emission line ratios in post-starbursts suggest that the H$\alpha$ fluxes are likely to be contaminated, and the D$_n(4000)$-based SFRs have greater uncertainties and trace a longer period of star formation, and are thus contaminated by the starburst itself. The SFRs derived from the extinction-corrected H$\alpha$ fluxes are 0.09, 0.58 M$_\sun$yr$^{-1}$ for H02 and S05 respectively. While the D$_n(4000)$-based SFRs have significantly higher uncertainties, the 68\%ile ranges are $0.01-0.29$ M$_\sun$yr$^{-1}$ for H02 and $0.006-1.05$ M$_\sun$yr$^{-1}$ for S05, consistent with the H$\alpha$-based SFRs.

We also consider several other SFR indicators to account for the possibility that the Balmer decrement measurements underestimate the true dust obscuration. We use the VLA FIRST Survey \citep[Faint Images of the Radio Sky at Twenty centimeters][]{Becker1995} to study the 1.4 GHz radio continuum emission as a SFR tracer. 1.4 GHz emission is often used as an extinction-free tracer of SFRs \citep{Condon1992}, but LINERs are found to contribute to the 1.4 GHz emission \citep{deVries2007, Moric2010} with greater scatter than for H$\alpha$ \citep{Moric2010}. Neither post-starburst galaxy considered here is detected in FIRST, and upper limits imply 1$\sigma$ upper limits on the SFRs of $<0.6$ and $0.5$ M$_\sun$yr$^{-1}$ for H02 and S05, respectively. 

\citet{Smercina2018} fit the dust continuum to estimate SFRs from the total infrared (TIR) flux, as well as the other infrared tracers [NeII]+[NeIII] and [CII]. S05 has a TIR SFR consistent with the other measures from the optical and radio, at 0.71 M$_\sun$yr$^{-1}$, and a lower SFR from [NeII]+[NeIII] of $<0.09$ M$_\sun$yr$^{-1}$. H02 has a significantly higher TIR SFR, at 4.6 M$_\sun$yr$^{-1}$. However, \citet{Smercina2018} find that the TIR flux is affected by A stellar population heating of the dust, which enhances the SFR estimate by a factor of $\gtrsim 3-4\times$, consistent with TIR being a much longer duration SFR tracer. 

The TIR flux could be an even larger overestimate of the true SFR, as \citet{Hayward2014} find the TIR flux overestimates the true SFR by $\gtrsim30\times$ during the post-starburst phase. If AGN heating is important, it would also act to boost the TIR flux, causing this tracer to be an overestimate of the SFR. A TIR SFR for H02 of $\sim 1$ M$_\sun$yr$^{-1}$ with the correction for A star heating would be consistent with the SFR estimated from [CII] for H02 (1.3 M$_\sun$yr$^{-1}$), and the 3$\sigma$ upper limit on the 1.4 GHz non-detection (1.8 M$_\sun$yr$^{-1}$). 

Thus, the SFRs from various tracers are consistent with a range of 0.09 (H$\alpha$) - 1 (TIR) M$_\sun$yr$^{-1}$ for H02, and 0.58 M$_\sun$yr$^{-1}$ or less for S05. The source of the discrepant SFRs for H02 between the low H$\alpha$ and even D$_n(4000)$ - derived values and the high IR - derived values could be either (1) an underestimate of the extinction compared to what is measured using the Balmer decrement, or (2) an underestimate of the dust heating from sources other than star formation, or a combination of both. The dust masses for the two galaxies are similar ($log$M$_{\rm dust} = 6.86 \pm 0.90, 7.00 \pm 0.16$ for H02 and S05 respectively, \citealt{Smercina2018}), so if the extinction is much higher in H02, the geometry must be different. Given the uncertainties in accurately determining the dust heating contribution from AGN or A stars, we adopt a conservative upper limit of $<1$ M$_\sun$yr$^{-1}$ on the SFR of H02. We consider the effect of this uncertainty in the SFR of H02 in \S\ref{sec:results}.

\section{Results}
\label{sec:results}

\begin{figure*}
\includegraphics[width = 1\textwidth]{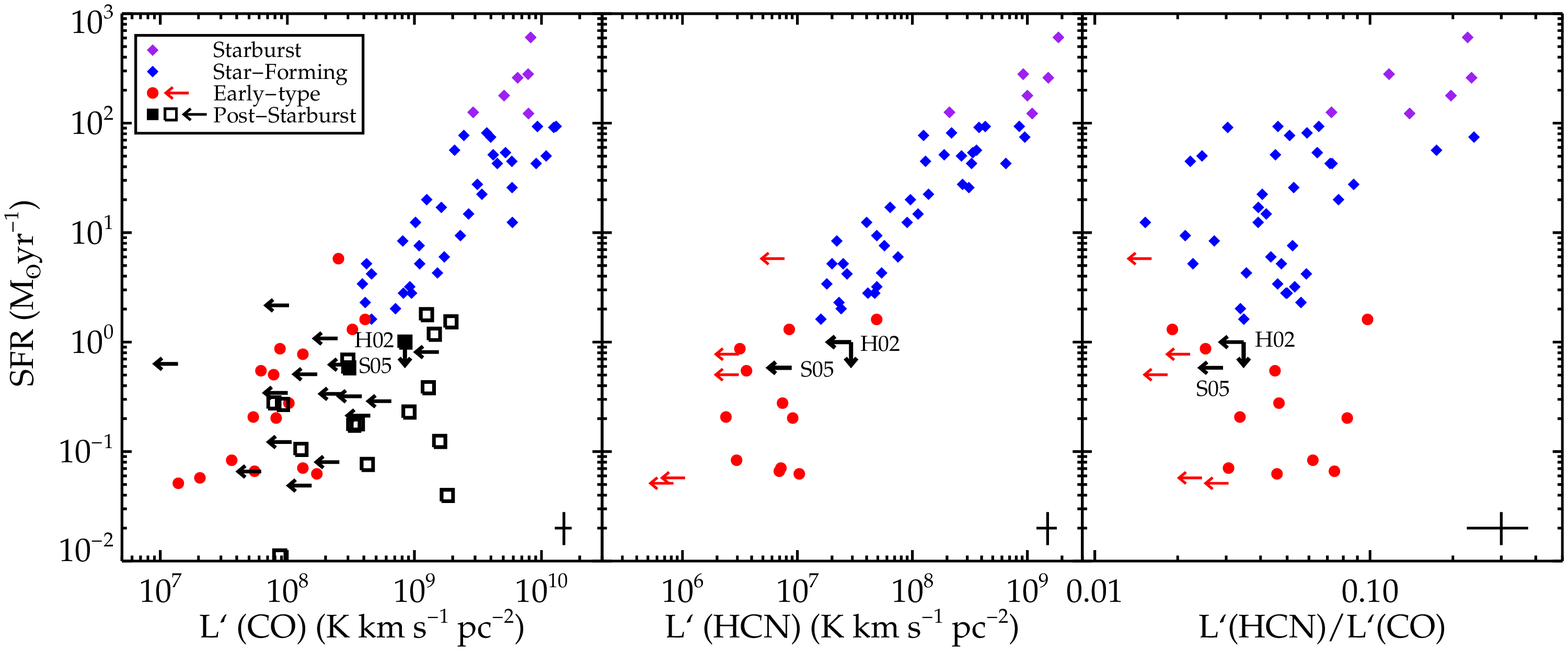}
\caption{{\bf Left:} SFR vs. $L^{\prime}($CO$)$ for star-forming and starbursting galaxies from \citet{Gao2004} (blue and purple diamonds), early type galaxies from \citet{Crocker2012} (red circles and arrows), and post-starburst galaxies \citet{French2015} (black squares and arrows). Filled black squares represent the two galaxies targeted for dense gas observations. Characteristic error bars are shown in the bottom right of each panel. All upper limits are at the $3\sigma$ level. The post-starburst galaxies have systematically low SFRs for their CO luminosities. The two post-starburst galaxies targeted for HCN observations are representative of the post-starburst population. {\bf Middle:}  SFR vs. $L^{\prime}($HCN$)$ for the same samples. HCN is not detected for either post-starburst galaxy studied here, consistent with their low SFRs and with the early type galaxies. The absence of denser gas traced by HCN reveals why the SFRs of post-starburst galaxies are so low. {\bf Right:} SFR vs. dense gas luminosity ratio  $L^{\prime}($HCN$)$/$L^{\prime}($CO$)$. The post-starburst galaxies targeted here have low HCN/CO luminosity ratios compared with the starbursting, star-forming and many CO-detected early type galaxies. The low HCN/CO luminosity ratios of the post-starbursts indicate the dense molecular gas fraction has changed since the starbursting phase and is different than in normal star-forming galaxies.}
\label{fig:hcn}
\end{figure*}

\begin{figure*}
\centering
\includegraphics[width = 0.7\textwidth]{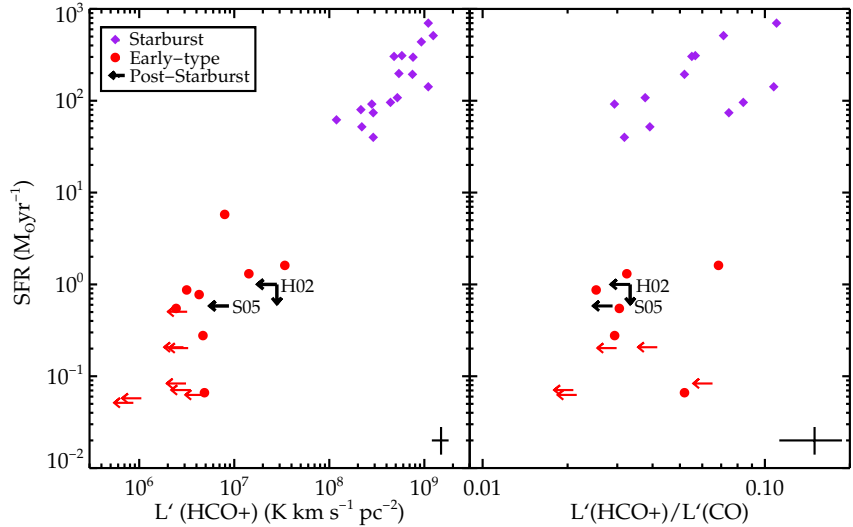}
\caption{{\bf Left:} SFR vs $L^{\prime}($HCO$^+)$ for starbursting galaxies from \citet{Gracia2008} (purple diamonds), early type galaxies from \citet{Crocker2012} (red circles and arrows), and post-starburst galaxies. All upper limits are at the $3\sigma$ level. HCO$^+$ is not detected for either CO-traced gas-rich post-starburst studied here, consistent with expectations from their low SFRs. {\bf Right:} SFR vs. dense gas luminosity ratio  $L^{\prime}($HCO$^+)$/$L^{\prime}($CO$)$. As with the HCN/CO luminosity ratio in Figure \ref{fig:hcn}, the post-starburst galaxies have low HCO$^+$/CO luminosity ratios compared with starbursting and many CO-detected early type galaxies. }
\label{fig:hco_plus}
\end{figure*}

We calculate the dense gas line luminosities using 
\begin{equation}
L^{\prime}_{line} = 3.25\times10^7 (1+z)^{-3} \ \nu_{\rm obs}^{-2} \ S_{line} \Delta v \ D_L^2
\end{equation}
where $L^{\prime}$ is the line luminosity in K\,km\,s$^{-1}$\,pc$^2$, $z$ is the redshift, $\nu_{\rm obs}$ is the observed line frequency in GHz, $S_{line}  \Delta v$ is the integrated flux density in Jy km/s, and $D_L$ is the luminosity distance in Mpc. We integrate over the velocity width of the CO (1--0) lines as measured using the IRAM 30m \citep{French2015}. In Table \ref{table:hcn}, we present the 3$\sigma$ upper limits on the HCN (1--0) and HCO$^+$ (1--0) line luminosities for the two post-starburst targets.

We compare the $L^{\prime}($HCN$)$ upper limits to the SFRs and CO line luminosities in Figure \ref{fig:hcn}. For comparison, we also show the rest of the \citet{French2015} post-starburst sample, as well as comparison samples of star-forming and early type galaxies with both HCN and CO measurements \citep{Gao2004, Crocker2012}. For the early type galaxies, SFRs are from \citet{Davis2014}, and we use the conversion factor from main beam temperature to flux density of 4.73 Jy/K from \citet{Young2011}. While the post-starbursts have high CO luminosities for their SFRs, the observed offset does not persist for the denser HCN-traced gas; their $L^{\prime}($HCN$)$ upper limits are consistent with their low SFRs. The $L^{\prime}($HCN$)$/$L^{\prime}($CO$)$ ratios for post-starbursts are low compared to the star-forming galaxies and most of the CO-detected early types. 

Similarly, we compare the HCO$^+$ upper limits to the SFRs for the post-starburst targets in Figure \ref{fig:hco_plus}. The comparison samples of star-forming and starbursting galaxies are from \citet{Gracia2008} and the early type galaxies from \citet{Crocker2012}. Again, the upper limits from the post-starburst targets are consistent with their quiescent SFRs, though the limits on H02 are higher than most of the quiescent comparison sample.

As discussed in \S\ref{sec:sfr}, the SFR for H02 is uncertain, with various tracers estimating $\sim 0.09-1$ M$_\sun$yr$^{-1}$. In Figures \ref{fig:hcn} and \ref{fig:hco_plus} we see that even at a SFR of $1$ M$_\sun$yr$^{-1}$, H02 still has a high $L^{\prime}($CO$)$ for its SFR, and yet a $L^{\prime}($HCN$)$ and $L^{\prime}($HCO$^+)$ consistent with the relation followed by the comparison galaxies.

In \citet{French2015}, we proposed several scenarios for explaining the discrepancy between the high CO luminosity and the low SFR aside from a low CO-traced star formation efficiency. Many are now disfavored given the consistency between the HCN luminosity and SFR. The observed discrepancy is not likely driven by un-accounted for dust extinction or significant low-mass star formation (a very bottom-heavy IMF) that is hidden from our SFR tracers. It is also improbable that significant star formation is missed in the aperture correction of the SFRs from the 3\arcsec\ (2.7-3.2 kpc) SDSS fibers (see discussion in \S\ref{sec:sfr}). The range of SFRs inferred from various tracers implies a large range of dust extinction values, and even so, the CO luminosities are systematically higher than expected from the low SFRs. Indeed, \citet{Smercina2018} find the SFRs are still much lower than expected given the high CO luminosities when extinction-insensitive IR-based SFR tracers are used.

The two post-starburst galaxies targeted here have HCN/CO luminosity ratios which are low compared to star-forming galaxies and many early types. A similar effect is seen in the HCO$^+$/CO luminosity ratios. Here, we compare the post-starburst galaxies to several different galaxy types: early type \citep{Crocker2012}, star-forming, and starbursting \citep{Gao2004}. We perform a Monte-Carlo analysis, drawing pairs of galaxies at random from these comparison samples, to test how unusual it is to find two galaxies with these HCN/CO luminosity ratios. Drawing from the combined sample of early type, star-forming, and starbursting galaxies, the probability of finding two galaxies with HCN/CO as low as the post-starbursts is 5\%.  Thus, the post-starbursts have relatively low HCN/CO luminosity ratios, and it is unlikely our observed upper limits are due to random selection. We also compare to the comparison galaxy populations individually, to address several specific questions.

Next we compare to the three galaxy types separately. The starbursting sample represents likely recent progenitors of H02 and S05. There are zero starbursting (SFR $\geq  100 $ M$_\sun$ yr$^{-1}$) galaxies with HCN/CO ratios as low as the post-starbursts. This large difference indicates the dense molecular gas fraction in the post-starburst galaxy has changed significantly since the starburst phase. Next, we consider a comparison to H02 and S05's likely descendants: the early type sample.  Drawing from just the early type sample, we find that two galaxies have HCN/CO luminosity ratios as low as the two post-starbursts 16\% of the time, thus, these observations alone are not sufficient to distinguish the post-starburst sample from the early types. The post-starburst sample may be already consistent with the early type galaxies, or a separate mechanism could be lowering the dense gas ratio in each (we discuss this further in \S\ref{sec:lowdensegas}).

So far, we have connected the lack of dense gas to why post-starburst galaxies are not forming stars. Here we explore why the HCN/CO luminosity ratio is low, so we compare to normal star forming galaxies (SFR $< 100$ M$_\sun$ yr$^{-1}$). While the post-starburst galaxies have CO luminosities similar to this comparison sample, the post-starburst SFRs are lower. Comparing the HCN/CO luminosity ratios, the post-starburst galaxies have a lower dense gas fraction. Drawing from the normal star-forming sample, we find two galaxies with HCN/CO as low as the post-starbursts only 2\% of the time in our Monte-Carlo analysis. This significant difference in dense molecular gas fractions implies the state of the molecular gas reservoir is substantially different than expected in the course of normal star-formation. We discuss several physical mechanisms which could alter the molecular gas state, lowering the dense gas fraction, in \S\ref{sec:lowdensegas}).

In summary, post-starburst galaxies have CO luminosities that are systematically high given their SFRs, when compared to the SFR-$L^{\prime}($CO$)$ relation that is followed by star-forming and early type galaxies. However, the low HCN and HCO$^+$ luminosities implied by our ALMA non-detections {\it are} consistent with the low SFRs in the two post-starbursts targeted here. These low dense gas luminosities and SFRs are typical of early type galaxies, the likely end-points of post-starburst evolution. The HCN/CO luminosity ratios are low compared to starbursting, star-forming and many CO-detected early type galaxies, implying a low fraction of dense molecular gas mass to total molecular gas mass.

\section{Discussion}
\label{sec:discussion}

\subsection{Interpretation of the $L^{\prime}($HCN$)$--SFR relation}
Despite the tight linear relation between HCN luminosity and SFR, there are still several uncertainties in the physical interpretation of this relation. Some have interpreted the fact that the $L^{\prime}($HCN$)$--SFR relation is more linear and has less scatter than the $L^{\prime}($CO$)$--SFR relation as evidence of a threshold density for star formation \citep{Wu2005, Heiderman2010, Lada2012}. However, the difference between these two relations can also be explained by whether the median gas density is above or below the critical density of CO (1--0) or HCN (1--0) \citep{Krumholz2007}, and the subsequent relation between the gas density and line luminosity given the emission from sub-thermal gas \citep{Narayanan2008}. Furthermore, \citet{Stephens2016} find that the kpc-scale observations of integrated galaxy properties cannot be explained by a simple summation of clumps, and suggest that the low scatter in the $L^{\prime}($HCN$)$--SFR relation is due to a universal dense gas star formation efficiency, universal stellar IMF, and universal core/clump mass functions, with the kpc scale being sufficient to sample the full mass functions as well as various evolutionary states. Thus, we do not necessarily expect the HCN traced gas to trace gas where collapse to stars is inevitable, and such a threshold is likely much higher than $10^4$ cm$^{-3}$ \citep{Krumholz2007b}.  Nevertheless, despite potential nuances in the interpretation of the linear SFR-HCN relations,  the HCN/CO ratio serves as a proxy for the dense gas fraction.

\subsection{Dense Gas in Other Post-Starburst-Like Galaxies}

There are two other post-starburst-like galaxies with dense gas observations in the literature. We do not include them in our analysis because they are not selected by our post-starburst selection criteria, but we discuss them here. The first is NGC 5195 (M51b). While the nucleus of this galaxy shows a post-starburst signature \citep[spectrum from][]{Heckman1980}, the integrated spectrum \citep{Kennicutt1992} does not have the significant Balmer absorption required to be selected into our sample. This galaxy was observed in CO (1--0) and HCN (1--0) by \citet{Kohno2002} with the Nobeyama 45m. Subsequent observations by \citet{Matsushita2010} did not detect HCN. Observations of NGC 5195 are complicated by the nearby spiral arm of M51 in this interacting system. Resolved measurements by \citet{Kohno2002} using the Nobeyama Millimeter Array (NMA) are brighter than the 15\arcsec\ beam unresolved observations. \citet{Alatalo2016a} reobserved this galaxy using the Combined Array for Research in Millimeter Astronomy (CARMA) and found a CO (1--0) line flux in between the two \citet{Kohno2002} measurements. NGC 5195 has both HCN {\it and} CO luminosities consistent with its SFR from \citet{Lanz2013}, unlike our targets. \citet{Alatalo2016a} conclude this galaxy has a star formation efficiency consistent with normal early type galaxies.

Another galaxy with some post-starburst characteristics and dense gas measurements is NGC 1266 \citep{Alatalo2014, Alatalo2015}. HCN (1--0) and CO (1--0) measurements are part of the Atlas-3D survey of early type galaxies \citep{Crocker2012}.  While this galaxy has nebular emission lines that would exclude it from our post-starburst sample, it is possible that this ``shocked" post-starburst galaxy is a precursor to our sample \citep{Alatalo2016}. NGC 1266 has the highest HCN/CO line luminosity ratio of any of the galaxies in the \citet{Crocker2012} sample, in contrast to the low dense gas ratios seen here. \citet{Alatalo2015} claim this difference may be caused by the molecular outflow in the nucleus of this galaxy enhancing the dense gas fraction in this region. NGC 1266 also has a unusually excited high-J CO emission, possibly from shocks \citep{Pellegrini2013}. In these integrated measurements, NGC 1266 appears to have a low SFR/$L^{\prime}($HCN$)$ and a normal SFR/$L^{\prime}($CO$)$. However, spatially resolved measurements show the galaxy to have a low SFR/$L^{\prime}($CO$)$ surface density ratio, perhaps driven by the galaxy's outskirts, with a normal SFR/$L^{\prime}($CO$)$ in its nucleus \citep{Alatalo2015}.

\subsection{Variations in luminosity to mass conversion}

In order to interpret the HCN and CO luminosities as tracers of molecular gas mass, we must use conversion factors to convert $L^{\prime}($HCN$)$ to a dense ($n \gtrsim 8\times10^{3}$ cm$^{-3}$) molecular gas mass upper limit and  $L^{\prime}($HCN$)$/$L^{\prime}($CO$)$ into the dense molecular gas mass to total molecular gas mass ratio (i.e., the ``dense gas mass fraction"). We assume that the conversion factor from $L^{\prime}($HCN$)$ to dense molecular gas mass is the same for post-starbursts and other galaxies. Therefore, our non-detections of $L^{\prime}($HCN$)$ imply low dense gas masses, relative to the dense gas mass --SFR relation traced by star forming and early type galaxies, explaining their quiescence.

We test whether the dense gas mass ratio could be normal, despite the observed dense gas luminosity ratio, due to uncertainties in the conversion factors.  What if $L^{\prime}($HCN$)$/$L^{\prime}($CO$)$ does not correlate with the dense gas mass fraction as it does for other galaxies? This scenario could occur if the CO luminosity to total molecular gas mass conversion factor $\alpha_{\rm CO}$ and HCN luminosity to dense molecular gas mass conversion factor $\alpha_{\rm HCN}$ vary differently with the state of the gas. 

A lower value of $\alpha_{\rm CO}$ is usually invoked in ULIRGs (Ultra Luminous Infrared Galaxies) and justified based on differences in the distribution or state of the gas, widening the linewidth \citep[e.g.,][]{Downes1998,Narayanan2012}. In \citet{French2015}, we consider whether a ULIRG-like $\alpha_{\rm CO}$ may be appropriate for post-starburst galaxies. While post-starbursts may be descendants of ULIRGs, we are observing them many dynamical times ($\sim10^6-10^{7.5}$ yr, \citealt{Genzel2010}) after the starburst phase has ended ($\sim$0.3-1 Gyr ago). We estimate the influence of the stellar potential on increasing the linewidth and lowering $\alpha_{\rm CO}$, and find it is not sufficient to resolve the observed offset between $L^{\prime}($CO$)$ and the low SFRs. Additionally, \citet{Smercina2018} find gas to dust ratios in post-starburst galaxies  consistent with nearby galaxies using a Milky-Way like value of $\alpha_{\rm CO}$, indicating significantly lower conversion factors are unlikely. The dense gas conversion factor is similarly uncertain, but to resolve the discrepancies for the post-starburst galaxies in both SFR-$L^{\prime}($CO$)$ and SFR-$L^{\prime}($HCN$)$,  some other effect would have to lower $\alpha_{\rm CO}$ without a decrease in $\alpha_{\rm HCN}$. 

We test this possibility using {\tt DESPOTIC} \citep{Krumholz2013} to model the change in $\alpha_{\rm CO}$ and $\alpha_{\rm HCN}$ for the typical Milky Way GMC and ULIRG conditions described by \citet{Krumholz2013}. ULIRG conditions result in $\alpha_{\rm CO}$ values $\sim5\times$ lower than in Milky Way GMC conditions, as expected. However, $\alpha_{\rm HCN}$
is lowered by the same factor. Thus, even ULIRG-like conditions could not generate the low observed HCN/CO luminosity ratios of post-starburst galaxies. The low HCN/CO luminosity ratios in the post-starburst targets are thus likely due to low dense gas mass ratios.

\subsection{What prevents the CO-traced gas from further collapse?}
\label{sec:lowdensegas}
The low observed upper limits on the HCN-traced gas present a puzzle when coupled with the observations of significant CO-traced gas: what prevents the CO-traced gas from further collapse to denser gas? In order to explain the low dense gas mass ratios, some mechanism must prevent collapse of the CO-traced gas, affecting the ability of dense gas to form. Here, we explore possible mechanisms to prevent this collapse.

One possibility for inhibiting collapse is if kinetic energy is injected into the gas, rendering it stable from gravitational collapse. An example of this process is ``morphological quenching" \citep{Martig2009, Martig2013}, where the increased shear in early type galaxies stabilizes the gas from gravitational collapse, decreasing the dense gas fraction and inhibiting star formation. We note that while the post-starbursts have a low dense gas fraction compared to the full set of comparison galaxies we consider, they may not have an especially low dense gas fraction compared to early type galaxies. If morphological quenching is already lowering the dense gas fraction in early type galaxies compared to star forming and starbursting galaxies, this would explain why we do not see a significant difference in the dense gas fractions of the post-starburst and early-type galaxies.

Another possible energy source is the dissipation of turbulence from AGN jets or shocks \citep{Nesvadba2010}. However, the role of turbulent energy is complex, even in ``normal'' star-forming galaxies, simultaneously increasing the dense gas fraction by driving collapse at small scales, and suppressing star formation by preventing the collapse of GMCs on larger scales \citep[e.g.,][]{Federrath2013}. \citet{Guillard2015} suggest that heating from the dissipation of AGN-injected turbulence inhibits gravitational collapse on all scales as it cascades.  

Conversely, turbulence from stellar feedback is invoked to explain the high gas densities and high dense gas fractions in ULIRGs \citep[e.g.,][]{Papadopoulos2012, Hopkins2013}. More detailed modeling of the gas state in this unusual suppressed state is needed to understand how various sources of kinetic energy can act to lower the dense gas fraction in these galaxies.

What then is the physical source of the injected energy?  Many dynamical times have passed since the starburst ended, so energy or turbulence from stellar feedback is unlikely. Secular sources of turbulence like morphological quenching act over $>1$ Gyr, so are unlikely to play a role during this short transitional phase.

We have selected the post-starburst galaxies on a lack of emission lines, which selects against strong AGN. Nevertheless, LINER-like activity is currently seen in many post-starburst galaxies \citep{Yan2006, Yang2006, French2015}. Obscured AGN activity in a small number of sources, including H02, is suggested by the dust spectral fitting done by \citet{Smercina2018}. Smercina et al. also observe H$_2$ rotational lines in S05 in the mid-IR, satisfying the turbulent heating threshold of H$_2$/7.7$\mu$m $>0.04$ seen in molecular hydrogen emission-line galaxies \citep[MOHEGs;][]{Ogle2010}. Further evidence of AGN activity affecting the molecular gas reservoirs is seen in the rapid decline of the CO-traced gas during the post-starburst phase, on a timescale too rapid to be explained by star formation (French et al. 2018 submitted to ApJ), with timescales and inferred outflow rates similar to those seen in LINERs \citep{Cicone2013}.

We also may be observing the post-starburst galaxies in the non-active phase of a trend of AGN variability. AGN activity is possible during the post-starburst phase \citep{Davies2007, Schawinski2009, Wild2010, Cales2015}, and varies with timescales short enough to observe turbulent gas without an active AGN. AGN are seen to turn on and off with timescales of $\sim10^4 - 10^5$ years \citep{Lintott2009, Keel2017}, and the timescale for turbulent energy from AGN to deplete is $\sim 10^7 - 10^8$ years \citep{Guillard2015}. One possible example of this is seen by \citet{Prieto2016}, who observe light echoes from past AGN activity in a post-starburst galaxy with a LINER-like center. AGN activity is also observed to suppress star formation in CO-traced gas \citep{Ho2005, Nesvadba2011, Guillard2015, Lanz2016}. 
Thus, while the details of how kinetic energy injected into the gas might lower the low dense gas fraction are still not understood, these post-starburst galaxies may have experienced recent AGN activity strong enough to disrupt the molecular gas and suppress star formation.

\subsection{Evolution to Early Type Galaxies}
The presence of large CO-traced molecular gas reservoirs in half of the post-starbursts studied also presents a puzzle in understanding how these post-starburst galaxies can evolve to normal early type galaxies. Post-starburst galaxies have stellar populations, color gradients, morphologies, and kinematics consistent with reaching the red sequence of early type galaxies in \citep{Norton2001, Yang2004, Yang2008, Pracy2013, Pawlik2015} in a few Gyr. However, early type galaxies are typically gas-poor, with molecular gas fractions of $\lesssim 10^{-3}$ \citep{Young2011}. What is the fate of the CO-traced gas reservoirs in these post-starburst galaxies? We find in French et al. (2018, submitted to ApJ) that the CO-traced molecular gas to stellar mass fraction declines with the time elapsed since the starburst ended, implying post-starburst galaxies should reach early-type levels of molecular gas in 700-1500 Myr. Thus, post-starbursts become gas poor as their stellar populations, color gradients, morphologies, and kinematics start to resemble early-types.

The CO-traced gas undergoes a dramatic transition as the galaxy evolves from starbursting to the post-starburst phase over $\lesssim$ 1 Gyr. Starburst galaxies have enhanced CO-traced SFEs compared to normal star-forming galaxies. After the starburst ends, we observe post-starburst galaxies to have suppressed CO-traced SFEs relative to normal star-forming galaxies. Over the $\sim$ Gyr of evolution between these two phases, the dense gas mass ratio also evolves from high to low, but the HCN luminosity tracks with the SFR throughout this process. Thus, despite the fact that collapse to star formation is not guaranteed at the $n \gtrsim 8\times10^3$ cm$^{-3}$ densities traced by HCN, the processes which drive the starburst, the end of the starburst, and the dramatic change in CO-traced SFEs do not affect the $L^{\prime}($HCN$)$ --SFR relation. This result is consistent with the idea proposed by \citet{Krumholz2007} and \citet{Stephens2016} that the dense gas SFE is universal on kpc scales. While these studies were based on star-forming and starbursting galaxies, our result suggests that this universality may extend to quiescent galaxies with low SFRs.

\section{Conclusions}

We survey the dense molecular gas content of two post-starburst galaxies possessing large reservoirs of CO-traced molecular gas, despite their lack of significant current star formation. ALMA does not detect either HCN (1--0) or HCO$^+$ (1--0) in these galaxies. This absence of denser gas is consistent with their low star formation rates. For the first time, we have direct evidence as to why post-starburst galaxies are now quiescent: the denser gas required for star formation is absent. The HCN/CO luminosity ratio upper limits are low compared to star-forming and many CO-detected early type galaxies, implying a low fraction of dense molecular gas mass to total molecular gas mass. 

The low HCN luminosities of the post-starburst galaxies are already consistent with the early type galaxies into which they are expected to evolve. However, the significant CO-traced gas and thus the low dense gas fraction necessitates a more detailed view of how these galaxies could evolve into gas-poor early types, and what prevents the CO-traced gas from collapsing further. The ($\sim$200 Myr) decline in the CO-traced molecular gas during the post-starburst phase (French et al. 2018 submitted to ApJ) is too rapid to be explained by star formation alone. Thus, any successful feedback model must predict that both the CO-traced gas declines over this rapid timescale and that the CO-traced gas is stable against collapse to denser gas, possibly via the same mechanism.

This picture of how star formation ends and the molecular gas reservoirs are depleted in galaxies undergoing rapid transition may be largely representative, as $\sim$40-100\% of galaxies are expected to evolve through this phase \citep{Zabludoff1996, Snyder2011,Wild2016}, and higher redshift post-starbursts are also observed to have large CO-traced molecular gas reservoirs \citep{Suess2017}.

%%%%%%%%%%%%%%%%%%%%%%%%%%%%%%%%%%%%%%%%%%%%%%%%%%%%%%%%%%%%%%%%%%%%%%

\acknowledgements

KDF is supported by Hubble Fellowship Grant HST-HF2-51391.001-A, provided by NASA through a grant from the Space Telescope Science Institute, which is operated by the Association of Universities for Research in Astronomy, Incorporated, under NASA contract NAS5-26555. AIZ acknowledges funding from NASA grant ADP-NNX10AE88G. DN acknowledges support from grants NSF AST-1724864, AST-1715206, and HST AR-13906,15043. AS acknowledges support for this work by the National Science Foundation Graduate Research Fellowship Program under grant No.~DGE 1256260. Any opinions, findings, and conclusions or recommendations expressed in this material are those of the author(s) and do not necessarily reflect the views of the National Science Foundation.

This paper makes use of the following ALMA data: ADS/JAO.ALMA\#2016.1.00881.S. ALMA is a partnership of ESO (representing its member states), NSF (USA) and NINS (Japan), together with NRC (Canada), NSC and ASIAA (Taiwan), and KASI (Republic of Korea), in cooperation with the Republic of Chile. The Joint ALMA Observatory is operated by ESO, AUI/NRAO and NAOJ. Funding for SDSS-III has been provided by the Alfred P. Sloan Foundation, the Participating Institutions, the National Science Foundation, and the U.S. Department of Energy Office of Science. The SDSS-III web site is http://www.sdss3.org/.

\bibliographystyle{apj}
\bibliography{earefs.bib}

\end{document}